\documentclass[10pt, twocolumn,
							 superscriptaddress,
							 english,
							 prb,
							 showpacs,
							 floatfix,
							 aps
							]{revtex4-2}
\usepackage{graphicx} 
\usepackage[utf8]{inputenc}
\usepackage{amsmath}
\usepackage{amssymb}
\usepackage{graphicx}
\usepackage{xspace}
\usepackage{xfrac}
\usepackage{xcolor}
\usepackage[backref=none,bookmarksnumbered=true,bookmarks=true,bookmarksopen=true,colorlinks=true,
citecolor=blue,linkcolor=blue,anchorcolor=green,urlcolor=blue,unicode=false]{hyperref}
\usepackage{ulem}

\newcommand{\twoH}{$2H$-NbSe$_2$}
\newcommand{\nbse}{NbSe$_2$}
\newcommand{\didv}{\ensuremath{\mathrm{d}I/\mathrm{d}V}\xspace}

\date{\today}

\begin{document}

\title{Odd-parity ground state in dilute Yu-Shiba-Rusinov dimers and chains}

\author{Lisa M. R\"{u}tten}
\affiliation{\mbox{Fachbereich Physik, Freie Universit\"at Berlin, 14195 Berlin, Germany}}

\author{Harald Schmid}
\affiliation{\mbox{Dahlem Center for Complex Quantum Systems and Fachbereich Physik, Freie Universit\"at Berlin, 14195 Berlin, Germany}}

\author{Werner M. J. van Weerdenburg}
\affiliation{\mbox{Fachbereich Physik, Freie Universit\"at Berlin, 14195 Berlin, Germany}}

\author{Eva Liebhaber}
\affiliation{\mbox{Fachbereich Physik, Freie Universit\"at Berlin, 14195 Berlin, Germany}}

\author{Kai Rossnagel}
\affiliation{\mbox{Institut für Experimentelle und Angewandte Physik, Christian-Albrechts-Universit\"at zu Kiel, 24098 Kiel, Germany}}
\affiliation{\mbox{Ruprecht Haensel Laboratory, Deutsches Elektronen-Synchrotron DESY, 22607 Hamburg, Germany}}

\author{Katharina J. Franke}
\affiliation{\mbox{Fachbereich Physik, Freie Universit\"at Berlin, 14195 Berlin, Germany}}

\begin{abstract}
      Magnetic adatoms on superconductors induce Yu-Shiba-Rusinov (YSR) states, which are key to the design of low-dimensional correlated systems and topological superconductivity. Competing magnetic interactions and superconducting pairing lead to a rich phase diagram. Using a scanning tunneling microscope (STM), we position Fe atoms on \twoH\ to build a dimer with an odd-parity ground state, i.e., a partially screened YSR channel with the hybridized states spanning the Fermi level. This ground state makes the dimer a promising precursor for a topological YSR chain. By adding one atom at a time, we track the formation of YSR bands. The lowest-energy band crosses the Fermi level and we find strong site-dependent spectral variations especially at the chain's terminations. We attribute these features to quantum spin effects and ferromagnetic coupling influenced by the local chemical environment, rather than topological superconductivity or Majorana modes.
\end{abstract}

\maketitle
\section{Introduction}
Spin chains on superconductors constitute fascinating systems for studying correlated one-dimensional systems and topological superconductivity \cite{Yazdani2023, Choi2019}. The magnetic adatoms are coupled to the superconductor by localized Yu–Shiba–Rusinov (YSR) states induced by exchange interactions \cite{Yu1965, Shiba1968, Rusinov1968}. Once two atoms are brought into sufficiently close distance, the YSR states hybridize, which can be described by symmetric and antisymmetric linear combinations of their YSR wave functions \cite{Ruby2018, Beck2021}. In the limit of long chains, the hybridized states evolve into YSR bands \cite{Schneider2021, Liebhaber2022, Kuester2022}. 

Depending on the exchange coupling strength to the superconductor, YSR states of single atoms can be in a free-spin or in a screened-spin ground state. In the latter case a quasiparticle is bound to the impurity. The transition between both regimes at a critical exchange coupling is associated with a quantum phase transition (QPT). Indirect magnetic coupling of neighboring adatom spins is possible in case of unscreened spins, where Rudermann-Kittel-Kasuya-Yosida (RKKY) \cite{Ruderman1954, Kasuya1956, Yosida1957} interactions are enabled by virtual excitations across the gap. Competing effects between quasiparticle screening and spin–spin interactions can drive quantum phase transitions in the adatom structures \cite{Yao2014a, Liebhaber2022, Steiner2021, Schmid2022}, altering the magnetic ground state. While it is difficult to directly probe the different magnetic ground states, indirect, yet distinct, fingerprints can be found in the excitation spectra within the superconducting gap. Mapping the excitation spectra while increasing the size of the adatom structure atom by atom thus, offers an intriguing route to the determination of the ground state of adatom chains \cite{Steiner2021}.

A particularly interesting situation occurs, if the YSR bands of an extended chain span the Fermi level, thus, leaving the chain in a partially screened ground state. It has been proposed that such a scenario is prone to show topological superconductivity and Majorana zero modes \cite{NadjPerge2013, Klinovaja2013, Pientka2013, Braunecker2013, Pientka2015}. The first theoretical models treated the spin in the classical limit as there was no experimental need to invoke the quantum spin nature. Indeed, the observations in densely packed chains were explained in the classical picture with an odd number of partially filled $d$ bands were crossing the Fermi level \cite{NadjPerge2013, NadjPerge2014, Ruby2015chains, Pawlak2016, Feldman2016, Ruby2017, Kim2018, Schneider2020, Schneider2022, Friedrich2021}. 

However, it was recently realized that at large adatom spacing, the quantum nature of the spins plays a significant role \cite{Liebhaber2022}, leading to a more complex phase diagram  \cite{Steiner2021}. For the design of topologically non-trivial chains the most promising avenue is to start with a dimer, where the two resonances related to one YSR component are found on both sides of the Fermi level. This is a regime where a YSR channel is partially screened and the ground state is of odd fermion parity. Upon forming a chain by adding atoms one expects partially filled YSR bands, which could potentially host topological superconductivity.

Using the tip of a scanning tunneling microscope, atoms can be manipulated and positioned with atomic site precision. This technique facilitates tuning of the interactions in YSR dimers and chains by varying the distance between the atoms and the lattice direction along which the atoms are arranged \cite{Kim2018, Schneider2021, Mier2021, Friedrich2021, Ding2021, Kuester2022, Liebhaber2022}. In addition, various substrate-adatom combinations can be investigated and spatially resolved spectroscopy allows the identification of states that only arise at the terminations of a chain as compared to states that persist throughout its bulk. Yet, the identification of end states is often further complicated by the small size of induced topological gaps alongside the limited energy resolution in experiments. These limitations also greatly hamper the observation of a hard topological gap on the bulk of the chain. Thus, experiments focused on ways to distinguish trivial from non-trivial end states for example by spin-polarized measurements \cite{Jeon2017}, or by their response to perturbations \cite{Schneider2020}. Other experiments tracked the opening of a small gap within the YSR band structure upon chain formation \cite{Schneider2022}.

Here, we position Fe atoms on the surface of a bulk \twoH\ crystal. Other than a previously investigated Fe chain on \nbse\ \cite{Liebhaber2022}, where the atoms were positioned parallel to an atomic Se row (i.e., the $[1\Bar{1}00]$ direction), we construct a dimer perpendicular to a Se row of the surface (i.e., the $[11\Bar{2}0]$ direction) and find it in an odd-parity ground state. As mentioned above, this ground state makes the dimer a promising start to realize a chain with a partially screened YSR band. We construct the corresponding chain atom by atom and find it to be partially screened. It further exhibits site-dependent spatial variations that are most pronounced at its terminations. However, we do not observe a clear gap opening in the bulk of the chain and spectral variations at the chain ends manifest as intensity shifts within the YSR band rather than well separated end states. From the partially screened ground state as well as differences in spectra recorded on the bulk of the chain compared to its terminations, we infer quantum spin effects and ferromagnetic coupling of the impurity spins.

\section{Experimental methods}
To enhance our energy resolution beyond the Fermi-Dirac limit, all data presented in this work were recorded using superconducting, Nb-coated tips. The tips were prepared by repeated indentation into an oxygen-reconstructed Nb crystal that was prepared by standard sputter and flash-anneal cycles. The indentations were performed until the spectra showed almost the full superconducting gap in the tip. The size of the tip gaps are indicated by gray shaded areas in the spectra and the precise values given in figure captions. Smaller tip crashes were performed to sharpen and stabilize the tip.

The chemical vapor transport-grown \twoH\ crystals were cleaved in ultra-high vacuum (UHV) using sticky tape (Tesa Kristallklar) and immediately transferred into the STM. Fe atoms were evaporated directly into the STM at temperatures below 12\,K. The Fe atoms were manipulated by laterally approaching the tip to an atom at a set point of a few nanoampere (1\,nA to 7\,nA depending on the tip apex) at 4\,mV. Each motion of the atom on the surface can be tracked by sudden jumps in the current and monitoring the $z$ traces while the atom is dragged across the surface by slowly moving the tip. When the desired position is reached, the tip is retracted to standard measuring set points. 

All differential conductance (\didv) maps shown in this manuscript were recorded in constant-contour mode, where the area of interest is first scanned in constant-current mode at the set point indicated in figure captions. The feedback loop is then opened, the bias voltage set to a value of interest, and the tip retraces the previously recorded constant-current trace while the \didv signal is recorded.

\section{Results}

Our main goal is the understanding of YSR hybridization and band formation in Fe chains along the $[11\Bar{2}0]$ direction of \twoH\ with a special focus on the states close to the Fermi level as these may develop into topologically non-trivial states. For a comprehensive picture, we start by describing the single Fe atom and dimer, although these have been already investigated in different contexts and regarding different aspects in earlier studies \cite{Liebhaber2020, Rutten2024}.
Our analysis of the Fe dimers goes beyond those presented in earlier works in that we not only observe a quantum-phase transition, but also deduce a level diagram of the corresponding YSR channel and find the new ground state to have odd fermion parity.

\subsection{Single Fe atoms: characteristic YSR patterns}

\begin{figure*}\centering	\includegraphics[width=\linewidth]{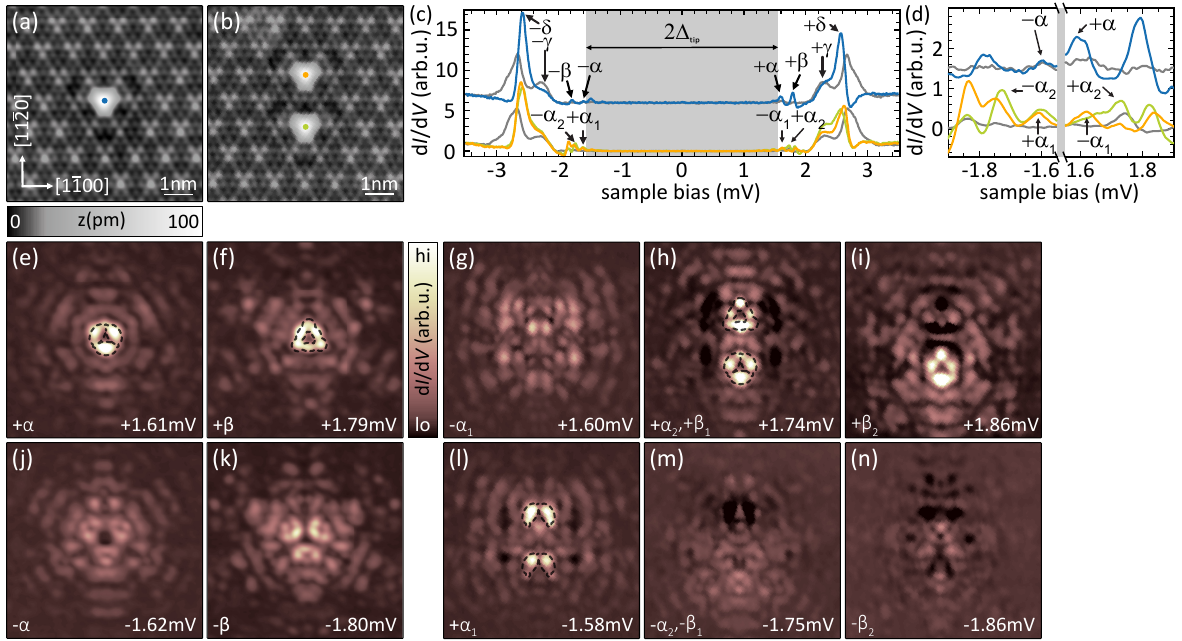}
	\caption{Fe monomer and dimer on the \twoH\ surface. (a), (b) Atomic resolution STM images of the monomer and dimer, respectively. (c) Tunneling spectra recorded on the Fe monomer and dimer adsorbed at CDW maxima of the \nbse\ substrate. For the color coding, refer to panels (a) and (b). Spectra of the \nbse\ substrate are shown in gray. (d) Zoom into the low energy region of (c). (e), (j) and (f), (k) \didv\ maps  of the monomer's $\pm\alpha$ and $\pm\beta$ resonances, respectively. The characteristic positive bias shapes are overlayed in black in (e), (f). (g)-(i), (l)-(n) \didv maps of the $\alpha$ and $\beta$-derived resonances of the dimer at both bias voltages.
	 $\Delta_{\mathrm{tip}}\approx$~1.55\,mV; set points: (a), (b) 10\,mV, 100\,pA; (e), (f), (j), (k), top row of (c) and (d): 5\,mV, 250\,pA; (g)-(i), (l)-(n), bottom row of (c) and (d): 5\,mV, 700\,pA. 
     }
	\label{fig:Fig1}
\end{figure*}

Figure\,\ref{fig:Fig1}a shows a STM topography of a single Fe atom on the \twoH\ surface. The atomic corrugation of the terminating Se lattice along with an additional $\sim3\times3$ modulation can be seen in the background. The additional modulation originates from the incommensurate charge-density wave (CDW), which coexists with superconductivity in \nbse\ at low temperatures. While the Fe atoms can reside in hollow sites or in metal sites (i.e., hollow sites of the Se lattice but with a Nb atom underneath) \cite{Liebhaber2020}, we focus on those in the hollow sites here. The YSR states of the Fe atoms in these sites have been shown to be strongly affected by the incommensurate charge-density modulations \cite{Liebhaber2020}. To minimize the effect of the CDW on our adatom structures, we position all atoms in maxima of the CDW. This adsorption site has also been used for the construction of other chains than presented in the following and for chiral structures \cite{Liebhaber2022, Rutten2024}. 

The differential conductance spectrum recorded centrally on the atom is shown in blue in Fig\,\ref{fig:Fig1}c. For comparison, a spectrum recorded on the bare substrate is shown in gray. We observe four YSR excitations that we label $\alpha$, $\beta$, $\gamma$, and $\delta$ from small to large energies as in previous works \cite{Liebhaber2020, Liebhaber2022, Rutten2024, Rutten2025}. The $\gamma$ and $\delta$ excitations coincide with the broad coherence peak structure of \twoH, which originates from its highly anisotropic band structure \cite{Sanna2022}. The four YSR resonances are caused by four singly occupied $d$ levels that are split by the crystal field. The $d$-level occupation leads to a
$S=2$ spin of the Fe adatom. Each of the $d$ orbitals can individually bind a quasiparticle on the superconductor \cite{Ruby2016, Choi2017}. The adatom spin is then (partially) screened or unscreened depending on the exchange coupling strength of the individual channels to the substrate. We denote the many-body states of the YSR system by its effective (i.e., unscreened) spin $S_\mathrm{eff}$ and the number of bound quasiparticles $Q$ \cite{Oppen2021}. In previous works the ground state of hollow-site Fe atoms at CDW maxima was found to most likely be $(Q=3, S_\mathrm{eff}=\sfrac{1}{2})$, where the $\alpha$, $\beta$, and $\gamma$ YSR channels are screened (i.e., bind a quasiparticle), while the $\delta$ channel is most likely unscreened \cite{Liebhaber2020}. In the following, we focus on the $\alpha$ and $\beta$ YSR states, which are both screened and located deep inside the superconducting gap. 

Each YSR state exhibits a characteristic spatial pattern given by the adsorption site on the crystal, the nature of the corresponding singly occupied $d$ level, and its coupling to the Fermi surface. Additionally, the electron and hole components of each YSR state are distinct from one another. Figure\,\ref{fig:Fig1}e and f show constant-contour \didv\ maps of the $+\alpha$ and the $+\beta$ ($+$ refers to the observation at positive bias voltage) resonance of the Fe monomer. For better recognition of the characteristic shapes we outline their contours by black dashed lines in Fig.\,\ref{fig:Fig1}e and f. The characteristic $+\alpha$ shape consists of three bright lobes; the characteristic $+\beta$ shape exhibits smaller lobes at the same positions as the $+\alpha$ shape complemented by three circles arranged in a triangle, which is rotated by 180$^\circ$ compared to the lobes. As a result, both characteristic shapes appear triangular but point in opposite direction. \didv\ maps of the negative bias voltage counterparts ($-\alpha$ and $-\beta$) are depicted in Fig.\,\ref{fig:Fig1}j and k. Note that these maps exhibit less intensity (same scaling for all maps) and different spatial patterns than those at positive bias. Because of these differences, we can use the characteristic shapes of the $+\alpha$ and the $+\beta$ resonance to relate resonances of Fe chains to their parent resonance in the monomer.

\subsection{Fe dimers along the $[11\Bar{2}0]$ direction of \twoH: odd-parity ground state}

We form a dimer oriented along the $[11\Bar{2}0]$ direction by positioning two Fe atoms in next-nearest neighbor CDW maxima. A topographic image of the resulting structure is shown in Fig.\,\ref{fig:Fig1}b. \didv\ spectra recorded centrally on both atoms are shown in the bottom row of Fig.\,\ref{fig:Fig1}c (see Fig.\,\ref{fig:Fig1}d for a close-up view on the low-energy region). Most notably, the number of YSR excitations in this energy region has increased compared to the monomer, indicating YSR hybridization. We label the observed states by $\pm\alpha_{1,2}$ and $\pm\beta_{1,2}$ according to their origin, as we will explain below. Note that the spectra on the two atoms differ from each other due to the absence of a mirror plane or inversion center between both atoms \cite{Rutten2024}. 

To gain further insights into the origin of the peaks in the \didv\ spectra, we plot the maps of the three lowest-energy features in Fig.\,\ref{fig:Fig1}g-i (positive bias) and Fig.\,\ref{fig:Fig1}l-n (negative bias). \didv\ maps of both components of the lowest energy YSR resonance are shown in Fig.\,\ref{fig:Fig1}g,l. Interestingly, we observe a shape that is closely resembling the $+\alpha$ shape (overlayed in black dashed lines), i.e., the shape that was found at positive bias voltage for the monomer, now at negative bias voltage. We also note that (contrary to the monomer and the higher-lying states) the map at negative bias voltage exhibits stronger intensity than its counterpart at positive bias voltage. The shape and reversed intensity indicate that one $\alpha$-derived state (we label it by $\alpha_1$) has crossed the quantum phase transition \cite{Farinacci2018}. We refer to the shape as reduced $+\alpha$ shape because one lobe is missing compared to the monomer's $+\alpha$ shape (Fig.\,\ref{fig:Fig1}l). This pattern can be understood as a result of the linear combination of the individual YSR wave functions, leading to destructive interference suppressing one of the lobes in the given geometry of the adatoms on the surface \cite{Rutten2024}. 

The maps of the next higher lying state are plotted in Fig.\,\ref{fig:Fig1}h and m. Here, we identify characteristics of both the monomer's $+\alpha$ and $+\beta$ shape at positive bias (Fig.\,\ref{fig:Fig1}h). The negative-bias resonance (Fig.\,\ref{fig:Fig1}m) remains with less intensity than the positive-bias counterpart. The observation of $+\alpha$ and $+\beta$ shapes at the same energy indicates that an $\alpha$- and a $\beta$-derived resonance overlap within our energy resolution. As we already identified one of the $\alpha$-derived states ($\alpha_1$) having crossed the quantum phase transition, it is worth emphasizing that the $+\alpha$-like resonance of the second $\alpha$-derived states (labeled $\alpha_2$) remains at positive bias. Thus, only one of the $\alpha$-derived states is found across the quantum phase transition. 
The third set of maps (Fig.\,\ref{fig:Fig1}i,n) only exhibits $\beta$-derived shapes with the characteristic $+\beta$ shape as well as higher intensity at positive bias voltage (Fig.\,\ref{fig:Fig1}i).

An important conclusion from the \didv\ spectra and maps is the presence of hybridized states, where the $+\alpha_1$ resonance crossed the Fermi level while the $+\alpha_2$ resonance remained at positive bias. The new ground state is thus partially screened in the $\alpha$ channel. 
The Fe atom then gains magnetic moment, which is available for substrate-mediated RKKY interactions among the two atoms. A similar observation of an RKKY-driven quantum phase transition has been made for Fe dimers in nearest-neighbor CDW maxima (i.e., along the $[1\Bar{1}00]$ direction of \nbse), where both $\alpha$ channels were found to become unscreened \cite{Liebhaber2022}.

\begin{figure}\centering	\includegraphics[width=\linewidth]{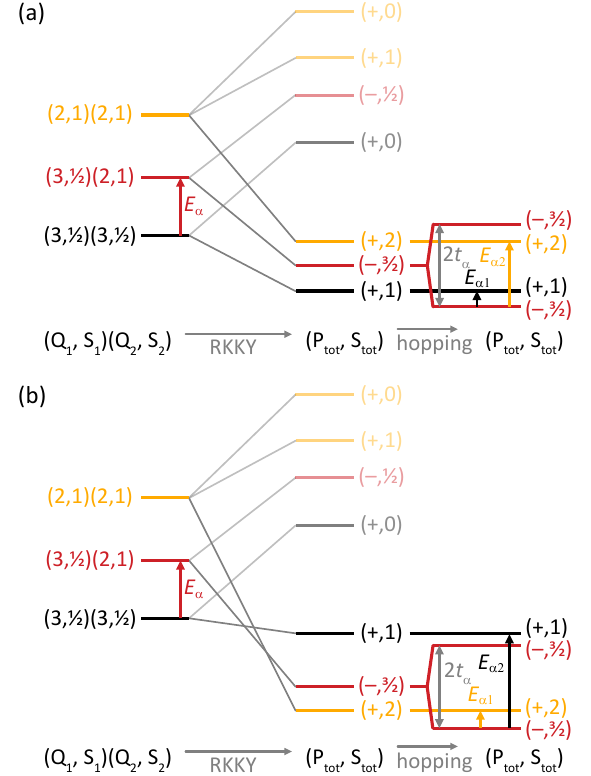}
	\caption{Schematic YSR level diagrams of the $\alpha$ channel of an initially non-interacting dimer and two scenarios (a), (b) of shifts as RKKY interaction and hopping between YSR states are successively switched on (left to right). Uncoupled monomers are labeled by individually screening quasi-particles $Q_{1,2}$ and effective spins $S_{1,2}$. The interacting dimer is labeled by its total parity $P_\mathrm{tot}$ and total spin $S_\mathrm{tot}$.
     }
	\label{fig:levels}
\end{figure}

Qualitative level diagrams in Fig.\,\ref{fig:levels} illustrate the competing interactions in the YSR system focusing on the $\alpha$ channel. To capture the effective spin-$\sfrac{1}{2}$ states of the monomer, we treat the dimer quantum mechanically. We assume that the $\alpha$ channel is independent from the other channels. We start with two non-interacting YSR monomers (left side). Here, we denote the states of the dimer by those of the individual monomers in the $(Q_i, S_i)$ notation, where $Q_i$ denotes the number of quasiparticles bound by impurity $i$ and $S_i$ denotes their resulting effective spin. The monomer ground state is $(3,\sfrac{1}{2})(3,\sfrac{1}{2})$, corresponding to three fully screened channels ($\alpha$, $\beta$, $\gamma$) and an unscreened $\delta$ channel. States related by the exchange $S_{1,\mathrm{eff}} \leftrightarrow  S_{2,\mathrm{eff}}$ (inversion symmetry) are degenerate. The selection rules for single electron tunneling are $\Delta Q = \pm 1$ and $\Delta S = \pm \sfrac{1}{2}$. We indicate the allowed transition by single-headed arrows. 

When RKKY interaction between the atoms is included, quantum numbers of the individual monomers are no longer conserved and we label states by the total fermion parity $P_\mathrm{tot}=\pm$ and the total spin $S_\mathrm{tot}$ of the dimer. All states with $S_\mathrm{i}>0$ on both atoms shift down in energy, where the shift is larger for higher total spins. Assuming ferromagnetic RKKY interaction (as will be justified later), the high spin states move lower in energy, while lower spin states (lighter colors in Fig.\,\ref{fig:levels}) shift up in energy (generally out of the superconducting gap \cite{Schmid2022}). 
Note that it is highly unlikely for the odd-parity state to become the ground state by RKKY interactions only \cite{Schmid2022} leaving us with two possible scenarios that are sketched in Fig.\,\ref{fig:levels}a and b. In Fig.\,\ref{fig:levels}a we plot a scenario in which the order of the levels remains unchanged by the RKKY interaction, but the spacing between the levels is reduced.  In the alternative scenario (similar to that observed for the dimer along the $[1\Bar{1}00]$ direction), the $(+,2)$ state becomes the new ground state due to the gain in RKKY energy overcompensating the costs for unscreening the $\alpha$ channel of both atoms \cite{Liebhaber2022}. This case is depicted in Fig.\,\ref{fig:levels}b.

Next, we also include hopping to our model which causes doubly degenerate levels to split (right side of Fig.\,\ref{fig:levels}). 
This hybridization splitting ($t_\alpha$, gray double headed arrows in Fig.\,\ref{fig:levels}) can exceed the energy difference between the RKKY induced ground state ($(+,1)$ in Fig.\,\ref{fig:levels}a and $(+,2)$ in Fig.\,\ref{fig:levels}b) and the $(-,\sfrac{3}{2})$ level. As a result, one of the hybridized $(-,\sfrac{3}{2})$ states becomes the new ground state. The quantum phase transition into a partially screened $\alpha$-derived dimer YSR channel is thus driven by a combination of RKKY coupling and hybridization splitting. 

Because the distinct difference between the two scenarios is the energetic order of the $(+,2)$ and $(+,1)$ excited states, one can distinguish them in experiment. In the first case (Fig.\,\ref{fig:levels}a), it is the small-energy excitation that has undergone the quantum phase transition. This is contrary to the second case (Fig.\,\ref{fig:levels}b), where it is the larger-energy excitation that reversed the states. In experiment, we identified the $\alpha_1$ with the smaller energy as the one having reversed intensity from positive to negative bias voltage. Hence, we assign the first scenario to our experimental observations. 

From experiment we also deduce a lower limit of the hybridization splitting as $t \geq E_{\alpha_1} \approx 40$\,$\mu$eV. This lower limit is comparable to the hybridization splitting found for the dimer along the $[1\Bar{1}00]$ direction in Ref.\,\cite{Liebhaber2022}, indicating at least comparable overlap of the YSR wave functions despite the larger spacing between the atoms along the $[11\bar20]$ direction.

The partial unscreening of an individual channel observed here is particularly interesting as it places the dimer in an ideal starting point for chains with the $\alpha$ YSR band crossing the Fermi level, which in turn is ideal for the development of a topologically non-trivial band structure.

\subsection{Short Fe chains along the $[11\Bar{2}0]$ direction: tracking YSR band formation}

\begin{figure*}\centering	\includegraphics[width=\linewidth]{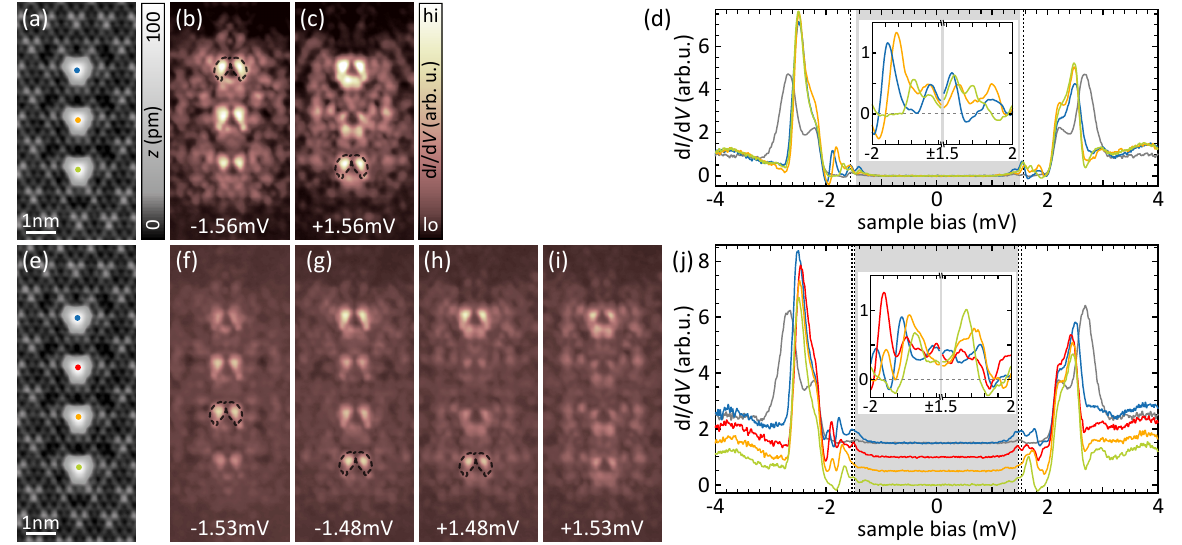}
	\caption{Fe\textsubscript{3} (top row) and Fe\textsubscript{4} (bottom row) chains on \twoH. (a), (e) Atomic resolution STM images of the trimer and tetramer. (b), (c) \didv maps of the trimer at both bias voltage polarities exhibiting characteristic $\alpha$ shapes (overlayed in black for exemplary sites). (d), (j) \didv spectra recorded centrally on each atom of the trimer (see (a) for color coding) and tetramer (see (e) for color coding), respectively. The spectra in (j) are offset along the \didv axis for clarity (not in the inset for better comparability). Vertical dashed lines indicate the energies at which the \didv\ maps in (b), (c), and (f)-(i) were recorded. Insets in (d) and (j) show close-ups on the low energy region with the region of tip gap omitted. (f)-(i) \didv maps of the tetramer exhibiting characteristic $\alpha$ shapes.
	 $\Delta_{\mathrm{tip}}\approx$~1.48\,mV; set points: (a), (e) 10\,mV, 50\,pA; (b)-(d), (f)-(j): 5\,mV, 750\,pA. 
     }
	\label{fig:short}
\end{figure*}

Having identified a dimer that is partially screened in one YSR channel, we continue to build a linear chain of Fe atoms by adding one atom at a time and refer to chains consisting of $n$ atoms as Fe\textsubscript{$n$} chains. Figure\,\ref{fig:short}a shows a topographic image of an Fe\textsubscript{3} chain. \didv spectra recorded on the individual atoms (Fig.\,\ref{fig:short}d) exhibit several overlapping resonances deep inside the superconducting gap indicating further hybridization in line with the observations on the dimer. In particular, we note that the states overlap with the Fermi level of the substrate, i.e., there is no gap around the energy of the tip gap (a zoom of the low energy region is shown in the inset of Fig.\,\ref{fig:short}d). The individual spectra differ from one another owing to the broken mirror symmetry along the chain. 
Although we cannot distinguish individual resonances in the spectra, we can obtain signatures of the origin of the hybridized states from \didv maps. The maps recorded at the energy of the peak deepest inside the gap (Fig.\,\ref{fig:short}b, c, energies indicated by dashed lines in Fig.\,\ref{fig:short}d) exhibit $+\alpha$ characteristics at both bias voltage polarities, indicating a ground state where the $\alpha$ YSR channel is partially screened.

We add another atom to form an Fe\textsubscript{4} chain (see Fig.\,\ref{fig:short}e for topography), and plot the corresponding \didv\ spectra in Fig.\,\ref{fig:short}j. As the expected number of YSR resonances increases with each additional atom, the larger spectral overlap makes the identification of individual peaks increasingly difficult. Notably, similarly to the trimer, the states overlap with the Fermi level. \didv\ maps recorded around the tip gap energy ($\pm\Delta_{\mathrm{tip}}\approx$~$\pm 1.48$\,mV) in Fig.\,\ref{fig:short}f-i reveal the characteristic $+\alpha$ shape. The broad states spanning the Fermi level indicate that the partially screened ground state is preserved with increasing chain length.

In analyzing the intensity distribution along the chain, we note that the characteristic $+\alpha$ shapes are more intense on the central atoms at higher energy (Fig.\,\ref{fig:short}f), whereas their intensity is stronger on the terminating atoms at smaller absolute bias voltage (Fig.\,\ref{fig:short}g). Both these trends persist as more atoms are added to the chain. Without loss of detail, we directly move on to an Fe\textsubscript{15} chain here. 

\subsection{Long Fe chains along the $[11\Bar{2}0]$ direction: edge effects in the YSR bands}

\begin{figure*}\centering	\includegraphics[width=\linewidth]{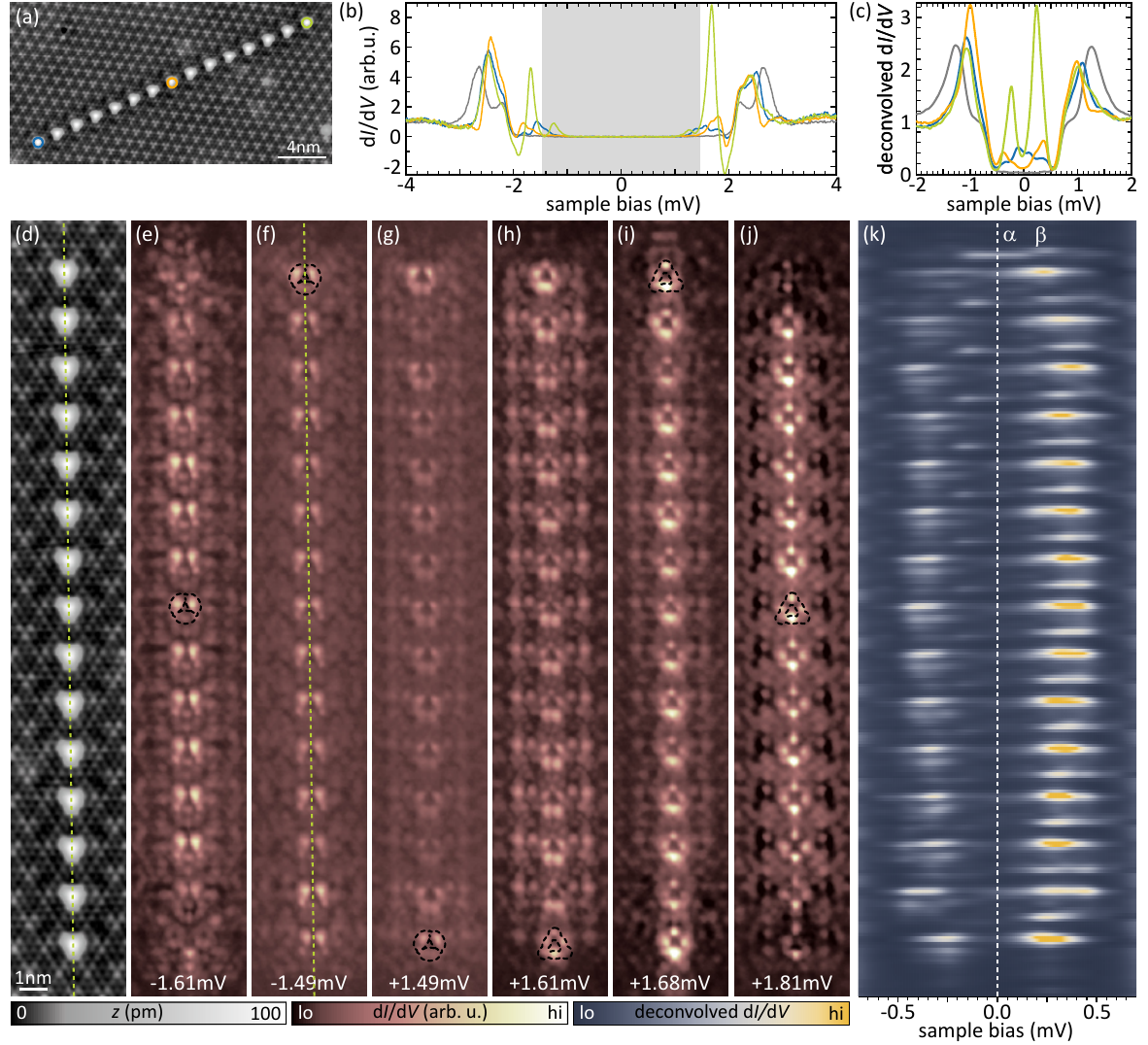}
	\caption{Fe\textsubscript{15} chain on \nbse. (a), (d) Atomic resolution STM images of the Fe\textsubscript{15} chain. (b) \didv spectra recorded on the central (orange trace) and both terminating (blue and green traces) atoms of the chain. (c) Same spectra as in (b) after numerically deconvolving the data. (e)-(j) \didv maps of the Fe\textsubscript{15} chain, where (e)-(h) show maps at both bias voltage polarities that exhibit mainly characteristics of the monomers $+\alpha$ resonance and (i), (j) show exemplary maps at positive bias voltage which exhibit $\beta$ characteristics. (k) False color contour plot plot of numerically deconvolved spectra recorded centrally along the chain (atom positions aligned with (d)-(j)).
    $\Delta_{\mathrm{tip}}\approx$~1.48\,mV; set points: (a), (d) 10\,mV, 50\,pA; (b)-(c), (e)-(k): 5\,mV, 750\,pA. 
     }
	\label{fig:long}
\end{figure*}

Figures\,\ref{fig:long}a and d show topographic images of the Fe\textsubscript{15} chain. \didv\ spectra recorded on the center and terminating atoms of the chain (see Fig.\,\ref{fig:long}b, positions where the spectra were taken are indicated by color coded circles in Fig.\,\ref{fig:long}a) exhibit broad features inside the superconducting gap, consistent with YSR band formation. Yet, the spectra vary depending on position. To gain a clearer understanding of the low-energy density of states along the chain, we numerically deconvolve the spectra and show the results in Fig.\,\ref{fig:long}c. While the intensity of the zero-energy excitation is higher on the terminating atoms of the chain compared to the central one, it remains nonzero in all spectra.

To obtain more detailed spatial information, we recorded spectra along the central axis of the chain (following the green dashed line in Fig.\,\ref{fig:long}d). The low-energy region ($\pm$0.70\,mV) of the numerically deconvolved spectra is presented as a false color plot in Fig.\,\ref{fig:long}k. The white dashed line marks zero excitation energy. Most notably, there is a clear distinction between the chain's termination and its bulk. 
At first glance, there appears to be a zero-energy excitation at one end of the chain (see top termination of the chain).
However, no corresponding counterpart is observed at the opposite end, excluding the presence of a Majorana zero mode. The inequivalence of the chain ends can once again be attributed to the chain's broken mirror symmetry. 

Contrary to the terminations, one may at first sight be tempted to identify a gap around zero energy within the chain in Fig.\,\ref{fig:long}k. However, closer inspection reveals a finite intensity all along the chain (as already mentioned in regard to Fig.\,\ref{fig:long}b,c). The corresponding intensity is mapped out in Fig.\,\ref{fig:long}f,g. The maps (most clearly the one at negative bias) reveal the reduced $\alpha$ shape (compare to the dimer and short chains in Fig.\,\ref{fig:Fig1}, Fig.\,\ref{fig:short}, respectively), reflecting the strong $\alpha$ character of the band spanning the Fermi level. Maps at slightly higher energy change character and signatures of the $\beta$ resonance add to the conductance (Fig.\,\ref{fig:long}h), suggesting the overlap of $\alpha$ and $\beta$ band. The $\beta$ character then becomes more pronounced at larger energies (Fig.\,\ref{fig:long},i,j). Most characteristically of the $\beta$ band is the additional intensity on the central axis of the chain. It originates from the small additional circles at the triangles' vertices of the $\beta$ state which persist upon hybridization (see dimer in  Fig.\,\ref{fig:Fig1}c,d). We outline the corresponding shape in some of the atoms in  Fig.\,\ref{fig:long}i,j. Turning back to the line spectra in Fig.\,\ref{fig:long}k, we now also understand from the characteristic shapes that the $\alpha$-derived band is generally of less intensity compared to the $\beta$-derived band on the central axis.

Having established the character of the bands, we stress again that the $\alpha$-derived band extends across the Fermi level. We have also seen local intensity variations of the band, most notably that there was higher intensity at zero energy at the chain's termination (Fig.\,\ref{fig:long}g), whereas the bulk of the chain shows stronger $\alpha$ shapes at larger energies (Fig.\,\ref{fig:long}e,h). Interestingly, we observe a similar trend of the energy-dependent intensity distribution in the $\beta$-derived band. At lower energies, most intensity is found on the terminating atoms (Fig.\,\ref{fig:long}i), while at higher energies, the intensity is larger in the bulk of the chain (Fig.\,\ref{fig:long}j).  

Next, we aim to understand the increase of the local density of states at the chain terminations in the $\alpha$- and $\beta$-derived YSR bands at low energy in contrast to the greater intensity observed in the bulk of the chain at larger energies. First, we note that the CDW remains in the same domain (hollow-centered) all along the chain. Therefore, we can exclude an influence of a varying potential on the YSR bands, which would lead to a  shift of the bands \cite{Liebhaber2022}. The effect of the CDW is seen in much longer chains in the Appendix. The spatial modulation of the local density of states, thus, arises intrinsically from the interactions within the chain. Second, we note that our chain is in the dilute limit. Electron hopping between the Fe sites is thus relatively weak, resulting in a narrow bandwidth of the YSR bands, consistent with the observed width of only $\sim 300$\,$\mu$eV. As the narrow bandwidth precludes the observation of individual resonances, we do not observe particle-in-a-box-like states as found in chains of densely spaced Mn atoms on Nb \cite{Schneider2021}. Instead, dilute chains are prime examples for correlated quantum spin systems, where the local excitation spectra crucially depend on the ground state and the magnetic coupling between the sites  \cite{Steiner2021}. 
The reason for the strong site dependence is the propagation of an excitation along the chain. For a fully screened chain such an excitation can propagate freely in a (mostly) spin-free background. For a completely unscreened chain an excitation (i.e., a screened site) can in principle propagate along the chain, but does so on a correlated spin background. As described above, the $\alpha$-derived band crosses the Fermi level and is in a partially screened regime, whereas the $\beta$-derived band is completely screened.  
The excitation spectra further depend on the nature of the magnetic coupling along the chain. In case of ferromagnetic coupling and considering only nearest-neighbor interactions, the excitation involves breaking of two ferromagnetic bonds within the chain, but only one at the chain ends. We thus expect higher intensity of the excitation at lower energy at the chain ends (similar to a simple tight-binding chain). On the contrary, anti-ferromagnetic coupling would not lead to a strongly confined modification of the excitation at the termination, but to a more extended spread of the spectral weight along the chain \cite{Steiner2021}. Our observation of a strong change in the excitation spectra in both the $\alpha$- and $\beta$-derived bands at the terminating atoms, therefore, indicates ferromagnetic coupling in the chain.


\section{Conclusions}
In summary, we assembled chains of Fe atoms along the $[11\Bar{2}0]$ crystallographic direction of \twoH. Hybridization and magnetic interactions drive the dimer into an odd-parity ground state, where the lowest-energy YSR channel is partially screened. This state is an ideal precursor for creating a partially-screened YSR band in longer chains.
As we extend the chain atom by atom, we track the progressive YSR hybridization and the formation of the resulting bands. Consistent with the expectations from the dimer, the extended YSR band in a chain of 15 Fe atoms spans the Fermi level. However, despite the promising electronic states, we do not find signatures of a topological gap or Majorana zero modes in the Fe\textsubscript{15} chain. Yet, we find strong variations of the YSR signal at the chain's terminations with one of the states appearing close to zero energy. These variations, however, can be attributed to trivial changes in the environment due to the absence of a neighboring atom. A simple model, in which the Fe atoms are coupled ferromagnetically via substrate-mediated RKKY interactions captures all key observations. While magnetic interactions in the quantum spin models are typically considered within individual bands, this approach may be too simplistic in our case as we find indications of overlapping YSR bands.

\acknowledgements


We thank Felix von Oppen and Gal Lemut for fruitful discussions. We gratefully acknowledge financial support by the Deutsche Forschungsgemeinschaft (DFG, German Research Foundation) through Projects No. 277101999 (CRC 183, Project No. C03) and No. 328545488 (CRC 227, Project No. B05). LMR acknowledges membership in the International Max Planck Research School ``Elementary Processes in Physical Chemistry".

\section*{Appendix: Effect of the CDW on long chains}

\begin{figure*}\centering	\includegraphics[width=\linewidth]{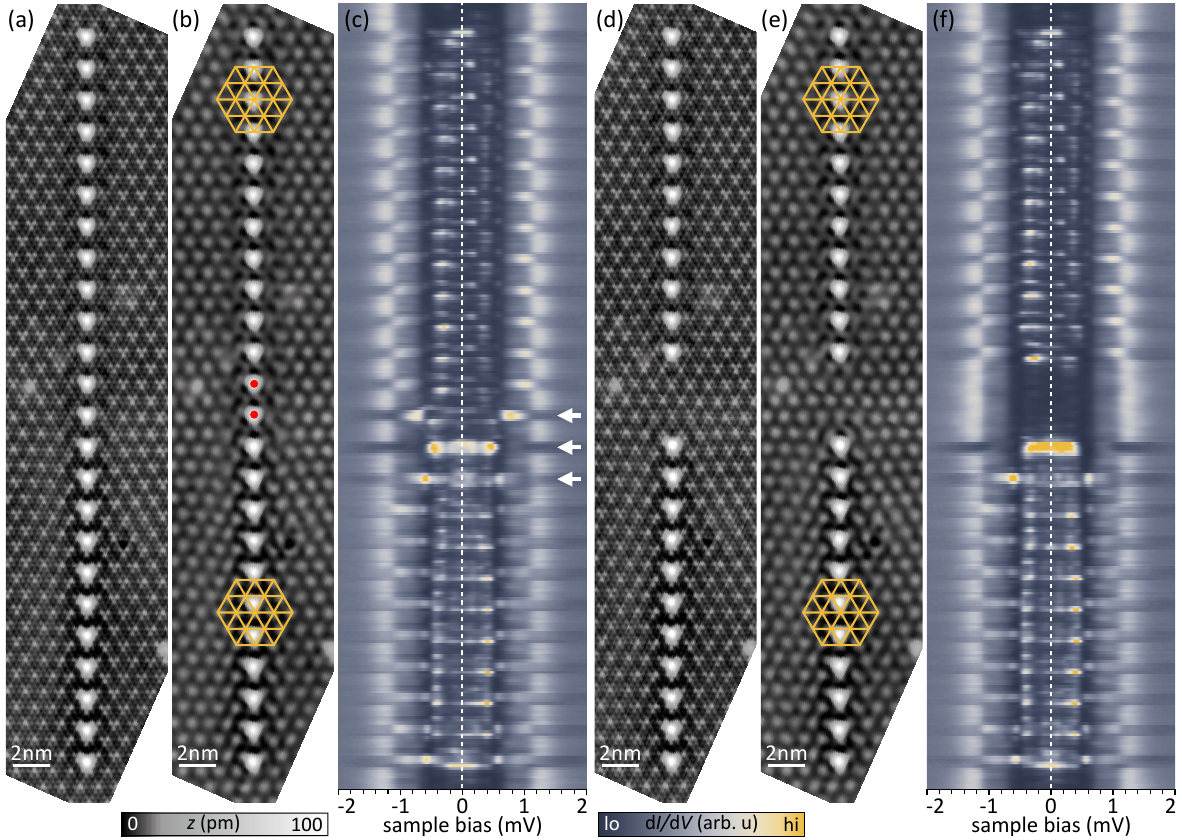}
	\caption{Fe\textsubscript{24} chain and two Fe\textsubscript{11} chains which were formed by removing the middle atoms from the Fe\textsubscript{24} chain on a varying CDW background. (a), (d) Atomic resolution STM images of the Fe\textsubscript{24} chain and the two Fe\textsubscript{11} chains, respectively. (b), (e) Same images as in (a), (d) after the atomic corrugation was removed by Fourier-filtering. Yellow grids visualize the CDW in two regions of the chains with CDW maxima located at their vertices. Red dots in (b) mark atoms that were removed to obtain two Fe\textsubscript{11} chains in (e). (c), (f) Contour plots of deconvolved \didv spectra recorded centrally along the Fe\textsubscript{24} chain and the two Fe\textsubscript{11} chains, respectively.
	 $\Delta_{\mathrm{tip}}\approx$~1.48\,mV; set points: (a), (b), (d), (e): 10\,mV, 50\,pA; (c), (f): 5\,mV, 750\,pA. 
     }
	\label{fig:SFigCDW}
\end{figure*}

If the chain is extended beyond Fe\textsubscript{15}, the atoms no longer all reside in equivalent CDW sites due to the incommensurability of the CDW. Figure\,\ref{fig:SFigCDW}a shows a topographic image of an Fe\textsubscript{24} chain, where variations in the CDW-to-lattice alignment can be observed around the chain. To reveal the positions of atoms with respect to the CDW along the chain we apply a Fourier filter to the topographic image to remove the atomic corrugation. The resulting image is shown in Fig.\,\ref{fig:SFigCDW}b. We exemplarily highlight different CDW positions of the Fe atoms towards both chain ends by overlaying yellow lattices where the crossings correspond to CDW maxima. These lattices reveal that the Fe atoms reside in CDW maxima in the upper part of the chain (as was the case for all chains discussed in the main text), while they sit in CDW minima in the bottom part of the chain. A false color contour plot of deconvolved spectra along the chain is depicted in Fig.\,\ref{fig:SFigCDW}c and reveals two regions in which resonances occur at constant energies, which we interpret as the formation of sub-chains. The two regions are connected by a three-atoms wide transition region (indicated by white arrows), where the spectra drastically differ from the rest of the chain. A similar behavior was observed for Fe chains along the $[1\Bar{1}00]$ direction of \twoH\ \cite{Liebhaber2022}. Because hollow-site Fe atoms at different positions with respect to the CDW exhibit different spectra, it is reasonable that YSR bands shift in energy as the atoms position with respect to the CDW changes due to the incommensurability of the CDW.

To test if the sub-chains are independent we remove the two central atoms of the Fe\textsubscript{24} chain (marked by red dots in Fig.\,\ref{fig:SFigCDW}b), thereby forming two Fe\textsubscript{11} chains. A topographic image of these chains and its Fourier-filtered version are shown in Fig.\,\ref{fig:SFigCDW}d and e. Note, that there are no obvious changes in the CDW-to-lattice alignment upon removal of the two atoms. Spectra recorded along the chain are plotted as a color plot in Fig.\,\ref{fig:SFigCDW}f and reproduce those shown in Fig.\,\ref{fig:SFigCDW}c everywhere except for the positions of the removed atoms and their nearest neighbors.


%

\end{document}